\RequirePackage{lineno}

\documentclass[12pt]{iopart}
\usepackage{hyperref}
\usepackage{graphicx}
\usepackage{upgreek}
\usepackage{gensymb}
\usepackage{amssymb}
\usepackage{textcomp}
\usepackage{setspace}

\begin{document}

\onehalfspacing

\title{Lattice dynamics of Coesite}

\author{Bj\"orn Wehinger$^1$, Alexe\"i Bosak$^1$, Aleksandr Chumakov$^1$, Alessandro Mirone$^1$, Bj\"orn Winkler$^2$, Leonid Dubrovinsky$^3$, Natalia Dubrovinskaia$^4$, Vadim Brazhkin$^5$, Tatiana Dyuzheva$^5$ and Michael Krisch$^1$}
\address{$^1$ European Synchrotron Radiation Facility, BP 220 F-38043 Grenoble Cedex, France}
\address{$^2$ Goethe Universit\"at Frankfurt, Institut Geowissenschaft, Abteilung Kristallographie, D-60438 Frankfurt, Germany}
\address{$^3$ Bayerisches Geoinstitut, Universit\"at Bayreuth, Universit\"atsstra\"se 30, D-95440 Bayreuth, Germany}
\address{$^4$ Material Physics and Technology at Extreme Conditions, Laboratory of
Crystallography, University of Bayreuth, D-95440 Bayreuth, Germany}
\address{$^5$ Institute for High Pressure Physics RAS, 142190 Troitsk Moscow region, Russia}

\ead{wehinger@esrf.fr}

\begin{abstract}
The lattice dynamics of coesite has been studied by a combination of diffuse x-ray scattering, inelastic x-ray scattering and an \textit{ab initio} lattice dynamics calculation. The combined technique gives access to the full lattice dynamics in harmonic description and thus eventually provides detailed information on the elastic properties, the stability and metastability of crystalline systems. The experimentally validated calculation was used for the investigation of eigenvectors, mode character and their influence on the density of vibrational states. High symmetry sections of the reciprocal space distribution of diffuse scattering and inelastic x-ray scattering spectra as well as the density of vibrational states and the dispersion relation are reported and compared to the calculation. A critical point at the zone boundary is found to contribute strongly to the main peak of the low energy part in the density of vibrational states. Comparison with the most abundant SiO$_2$ polymorph - $\alpha$-quartz -  reveals similarities and distinct differences in the low-energy vibrational properties.
\end{abstract}

\maketitle

\section{Introduction}
Coesite, SiO$_2$, is the highest density tetrahedrally coordinated crystalline polymorph of silica with space group C 1 2/c 1 \cite{gibbs_zkri_2003}. It was first synthesized in 1953 at 3.5 GPa and 750 $^{\circ}$C \cite{coes_science_1953} and later found in sandstone of the Arizona Barringer crater \cite{chao_science_1960} and led to the general acceptance of the impact cratering theory and to important implications for the recognition of meteorite impact craters in quartz-bearing geologic formations \cite{mark_UAP_1995}. The discovery of coesite revolutionized the whole meteorite study and the analysis of impact products. It was also shown that being encapsulated in diamond formed in deep Earth''s interior, coesite allows the unambiguous identification of `fossilized' high pressure states in individual inclusions of mantle samples delivered to the Earth's surface in the course of various geological processes \cite{sobolev_pnas_2000}. The fossilized pressure phases of the inclusion and the thermoelasticity of the host-inclusion (diamond-coesite) ensemble may provide a highly accurate geobarometer \cite{sobolev_pnas_2000}. Elastic and dynamical properties of coesite have been intensively studied in the past. Elastic constants for instance were examined using Brillouin light scattering \cite{weidner_jgr_1977} and ab initio calculations \cite{kimizuta_jap_2008}; vibrational properties at zero momentum transfer were measured by Raman \cite{liu_pcm_1997,sharma_nat_1981} and infrared spectroscopy \cite{lippincott_jrnbs_1958, williams_jgr_1993}. The compression mechanism was studied by single crystal x-ray diffraction \cite{angel_pcm_2003,levien_amin_1981} and ab initio calculations \cite{kimizuta_jap_2008}. The investigation of silica under pressures lead to the discovery of unexpected phenomena, such as, for example, the formation of mesoporous coesite at a pressure of 12 GPa and a temperature of 300 $^{\circ}$C \cite{mohanty_ac_2010,mohanty_jac_2009}.
Currently, the study of the phonon dispersion relations is limited to numerical calculations using potential based methods \cite{dean_prb_2000}. Experimental studies of dispersion relations have not been presented yet. The knowledge of the full lattice dynamics is, however, fundamental for the understanding of the compression mechanism and phase transitions. In particular the origin of the first peak of the density of vibrational states (VDOS) is of interest regarding the origin of the Boson peak in silica glass. As suggested in Ref. \cite{chumakov_prl_2011} the Boson peak in glasses originates from the acoustic phonon branches near the boundary of the pseudo-Brillouin-zone and has its counterpart in the VDOS of the corresponding crystal.
In the following we report the results of the powerful combination of diffuse scattering and inelastic x-ray scattering (IXS) for the study of distinct lattice dynamical features in a twinned crystal and the validation of the DFPT calculation. The combined approach allows understanding the lattice dynamics at arbitrary momentum transfers over the entire energy range and thus delivers a complete picture. The validated calculation is used for the study of particularities in the dispersion relations and contributions of different modes to the VDOS. 
The paper is structured as follows. The synthesis of the high quality crystal and the details of the measurement are given in Experimental methods. The lattice dynamics calculation is described in section Calculation. Experimental and theoretical results are presented and discussed in Results and Discussion. The Conclusion summarizes our findings.  

\section{Experimental methods}
\label{sec:exp}

\begin{figure}[ht]
\includegraphics[width=0.6\textwidth]{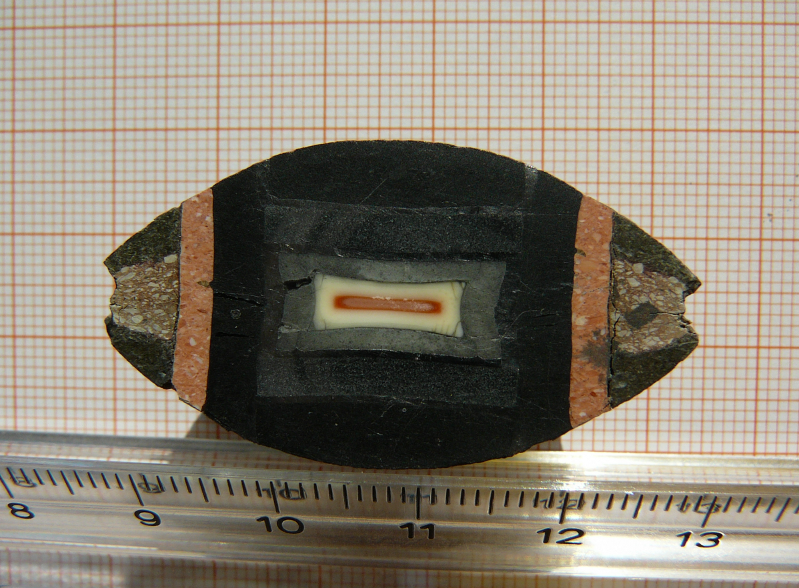}
\caption{\label{fig:sample} (colour online) Cross-section of the pressure pellet containing a polycrystalline sample of coesite synthesised at P= 5.5 GPa and T=1000 °C.}
\end{figure}

A polycrystalline sample of coesite was synthesized at the university of Bayreuth using the high-pressure high-temperature technique at P = 5.5 GPa and T = 1000 $^{\circ}$C. A 1000-ton hydraulic press (Voggenreiter GmbH) with a toroidal-type high-pressure cell \cite{khvostantsev_hthp_1977} was employed. The size of the synthesized sample was of about 50-70 mm$^3$ of pure coesite. As a starting material compacted amorphous silica (Sigma Aldrich, 99.9 \% purity) was used. The cross-section of the sample container just after the synthesis is shown in Figure \ref{fig:sample}. The single crystals were grown by the hydrothermal method described elsewhere \cite{dyuzheva_cryr_1998}.
The x-ray diffuse scattering experiment was conducted on beamline ID29 \cite{deSanctis_jsr_2012} at the European Synchrotron Radiation Facility (ESRF). Monochromatic X-rays with wavelength 0.7 \AA \hspace{1pt} were scattered from an elongated ($\approx 0.1\times 0.3 $ mm$^2$) coesite crystal at room temperature in transmission geometry. The sample was rotated with an increment of 0.1$^{\circ}$ orthogonal to the beam direction over an angular range of 360$^{\circ}$ while diffuse scattering patterns were recorded in shutterless mode with a PILATUS 6M detector \cite{kraft_jsr_2009}. The orientation matrix and geometry of the experiment were refined using the CrysAlis \cite{ox_diff} software package. The sample was found to be twinned in the $ac$ plane. 
The single crystal IXS study was carried out on beamline ID28 at the ESRF. The spectrometer was operated at 17.794 keV incident energy, providing an energy resolution of 3.0 meV full-width-half-maximum with a beam focus of 30 $\times$ 60 $\mu$m. IXS scans were performed in transmission geometry along selected directions in reciprocal space. Further details of the experimental set-up and the data treatment can be found elsewhere \cite{kirsch_Springer_2007}. 

The generalized x-ray weighted VDOS (X-VDOS) was obtained from IXS spectra of a polycrystalline sample measured at ID28 \cite{bosak_prb_2007}. The scattered radiation was collected by nine crystal analysers. The momentum transfer resolution of each analyser was $\approx$ 0.3 nm$^{-1}$.  The values of the momentum transfers for each analyser were chosen away from the Debye-Scherrer rings and covered the [10 : 70] nm$^{-1}$ range. The data combine the results of measurements with 1.4 meV resolution at 23.725 keV incident energy within [-25 : +25] meV energy range and 0.2 meV energy steps and results from the measurement with 3.0 meV resolution within [-25 : +180] meV energy range and 0.7 meV steps. The elastic peak in the IXS spectra was subtracted using the instrumental function of each analyser determined by x-ray scattering from a polymethylmethacrylate (PMMA) sample close to the maximum of its structure factor. The X-VDOS was obtained from the summed IXS spectra within the incoherent approximation following the data treatment procedure established in \cite{bosak_prb_2007}. 

\section{Calculation}
The lattice dynamics calculation was performed using the DFPT approach \cite{gonze_prb_1997} as implemented in the CASTEP code \cite{clark_zkri_2005,refson_prb_2006}. The local density approximation within the plane-wave pseudopotential formalism was employed using norm conserving pseudopotentials. The plane wave cut-off was set to 800 eV and the electronic structure was computed using a $3 \times 3 \times 2$ Monkhorst-Pack grid. The structure was optimized with the Broyden-Fletcher-Goldfarb-Shannon method, which is a full geometry optimisation of lattice and internal parameters. The cell parameters of the optimized cell are compared to the experimental values \cite{smyth_jpc_1987} from single-crystal neutron and x-ray diffraction in Table \ref{tab:cell}. They agree within 1.6 \%.

Phonon frequencies and eigenvectors were computed on a $4 \times 4 \times 3$ Monkhorst-Pack grid of the irreducible part of the Brillouin zone by a perturbation calculation and further Fourier interpolated for the VDOS and dispersion relations. The calculated phonon energies were tested to be converged to $<$ 0.05 meV.
Thermal diffuse scattering (TDS) and IXS intensities were calculated from the phonon eigenvectors and frequencies following the previously established formalism \cite{kirsch_Springer_2007, xu_zkri_2005, bosak_zkri_2012}, assuming  the validity of both the harmonic and adiabatic approximation. TDS intensities were calculated in first order approximation.

\section{Results and discussion}
\label{sec:results}

\begin{figure}[tb]
\includegraphics[width=1.0\textwidth]{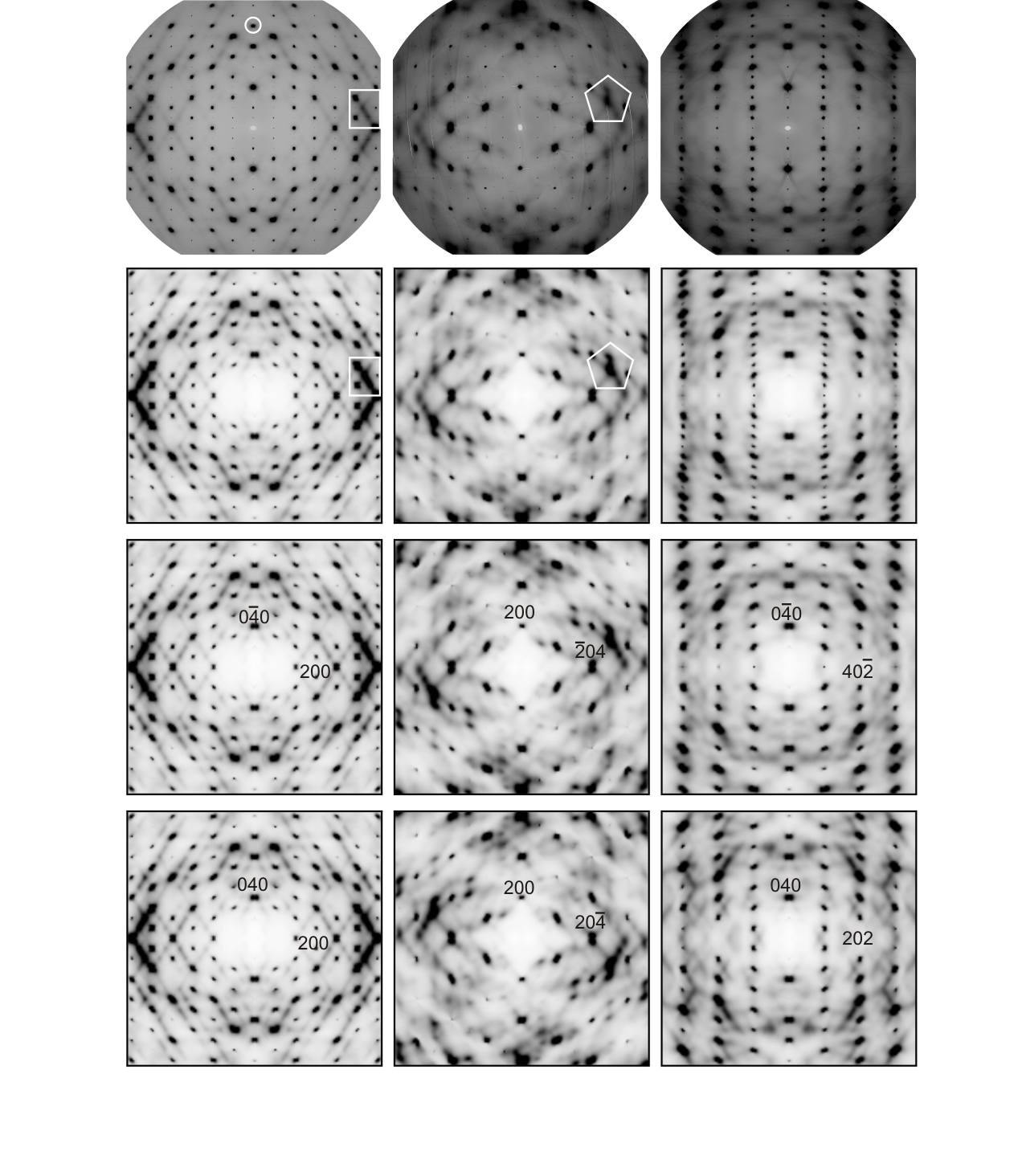}
\caption{\label{fig:TDS} Experimental (first row) and calculated (second row) diffuse scattering intensity distributions of coesite in high-symmetry reciprocal space sections. The absolute intensity is scaled to the best visualisation of the diffuse features. The calculated intensity distributions in the second row are created by the weighted superposition of the two crystalline orientations shown in the third and fourth row. The diffuse features marked by circle and polygon were selected for an IXS study, the intense features highlighted  by rectangle and polygon are guides for the comparison of experimental and calculated diffuse scattering. See text for further details.}
\end{figure}

High symmetry reciprocal space sections of diffuse scattering as obtained from experiment and corresponding calculated TDS intensity distributions are shown in Figure \ref{fig:TDS}. Corrections for polarisation and projection \cite{he_Wiley_2009} and the Laue symmetry of the system were applied. A complex intensity distribution is noticeable. Intense features indicated in Figure \ref{fig:TDS} were selected for a detailed IXS measurement in order to distinguish between possible elastic and inelastic contributions and to resolve the energies of the phonons contributing most to the TDS. A remarkable consistence between the experimental and calculated patterns can be seen for instance from the shape of indicated features. The twinning is mostly visible in the third column in Figure \ref{fig:TDS}. The number of Bragg reflections can only be described by a combination of two domains. The HK0 plane is common. The structure could be solved for a twinning of 180$^{\circ}$ around the reciprocal lattice axis a* with an intensity contribution of 25\% of the smaller crystal. The corresponding calculated TDS maps in the second row was created by the weighted superposition of the indicated intensity distributions. Experimental artefacts due to non-uniform absorption of the sample are visible, arising from the anisotropic shape of the sample. 

\begin{figure}[tb]
\includegraphics[width=0.6\textwidth]{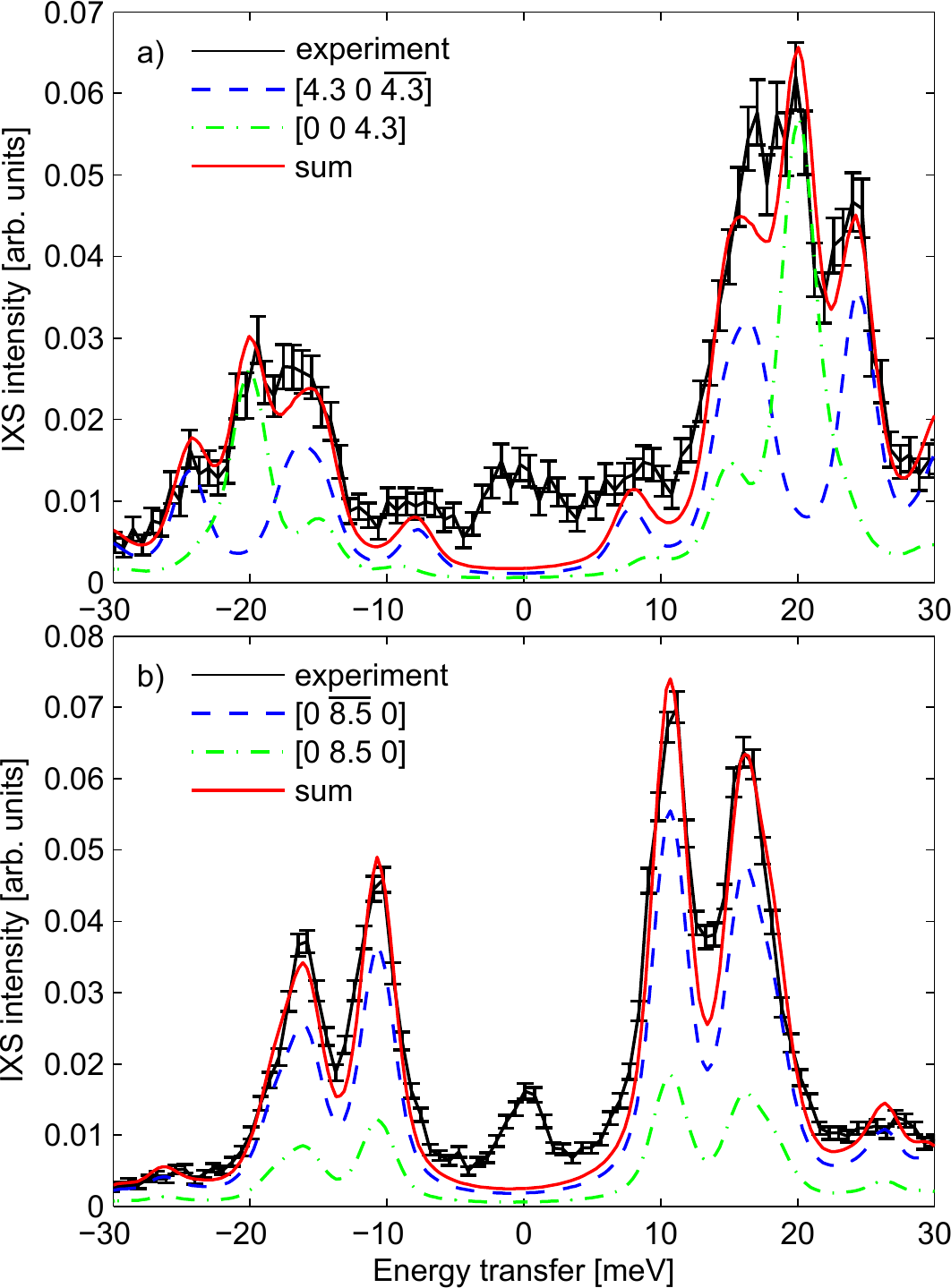}
\caption{\label{fig:IXS_spectra} (colour online) Experimental (black points with errorbars) and theoretical (red line) IXS spectra from coesite crystal. The theoretical spectra are the weighted sum of the two crystalline orientations: domain 1 - 75\% (blue dashed line), domain 2 - 25\% (green dotted dashed line).  Theoretical intensities were convoluted with the experimental resolution function and the energy transfer was scaled by 1.045.}
\end{figure}

\begin{figure}[tb]
\includegraphics[width=1.0\textwidth]{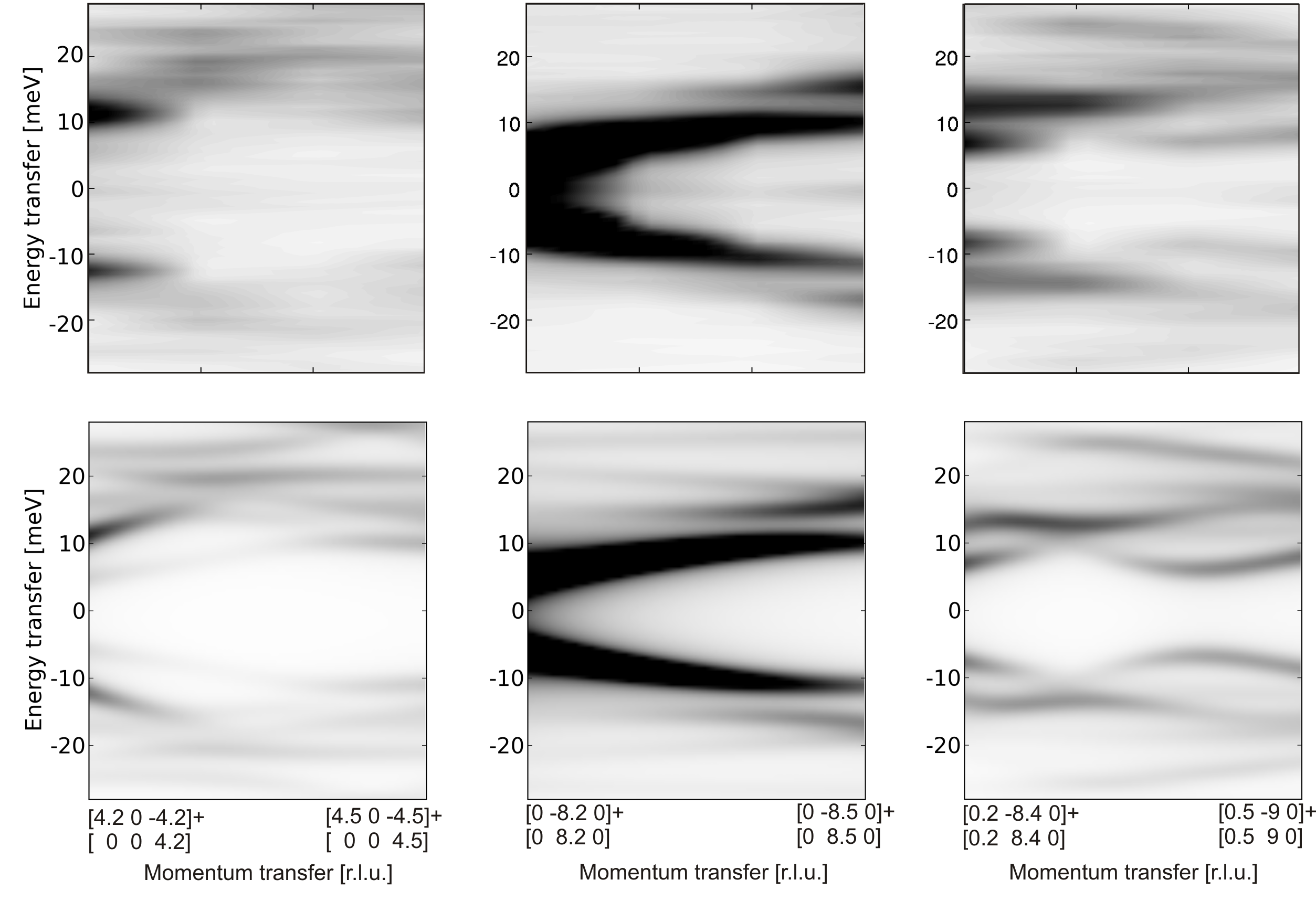}
\caption{\label{fig:IXS_maps} Experimental IXS intensity maps (first row) from coesite crystal together with theoretical intensity maps (second row) along the indicated directions. See text for further details.}
\end{figure}

Figure \ref{fig:IXS_spectra} shows two representative IXS spectra. One can clearly note the influence of the two crystalline domains on the inelastic spectra in Fig. \ref{fig:IXS_spectra} a). The spectra of the two domains belong to different $Q$-values with different phonon eigenvectors and energies, and are thus different. The calculated spectra of the two domains in the common plane \ref{fig:IXS_spectra} b) are, except for the overall intensity, identical. The experimental spectra also show a small elastic line. After scaling the calculated energies by 1.045 the theoretical spectra reproduce quite well both position and intensity of the phonons. The scaling factor was determined from the VDOS, and its value is justified further below. 
Figure \ref{fig:IXS_maps} shows the IXS intensity maps along the indicated directions from calculation and measurement. Each experimental map consists of 4 measured spectra with linear Q-spacing and energy steps of 0.7 meV. The momentum- and energy- transfers are linearly interpolated to 20 q-points and 72 energy steps. The theoretical maps were obtained by the weighted superposition of inelastic spectra for the two domains. The inelastic intensity is calculated from the eigenvectors and eigenfrequencies for 120 points along the given direction in reciprocal space and convoluted with the experimental resolution function of the spectrometer. The experimental spectra show that the diffuse scattering is of almost exclusively inelastic nature. Taking into account the intensity contribution of the two crystalline domains, both, energy transfer and inelastic intensities of the different phonon branches are well reproduced by the calculation for arbitrary directions. This implies that the theory correctly predicts both eigenvalues and eigenvectors at arbitrary momentum transfers.

\begin{figure}[tb]
\includegraphics[width=1.0\textwidth]{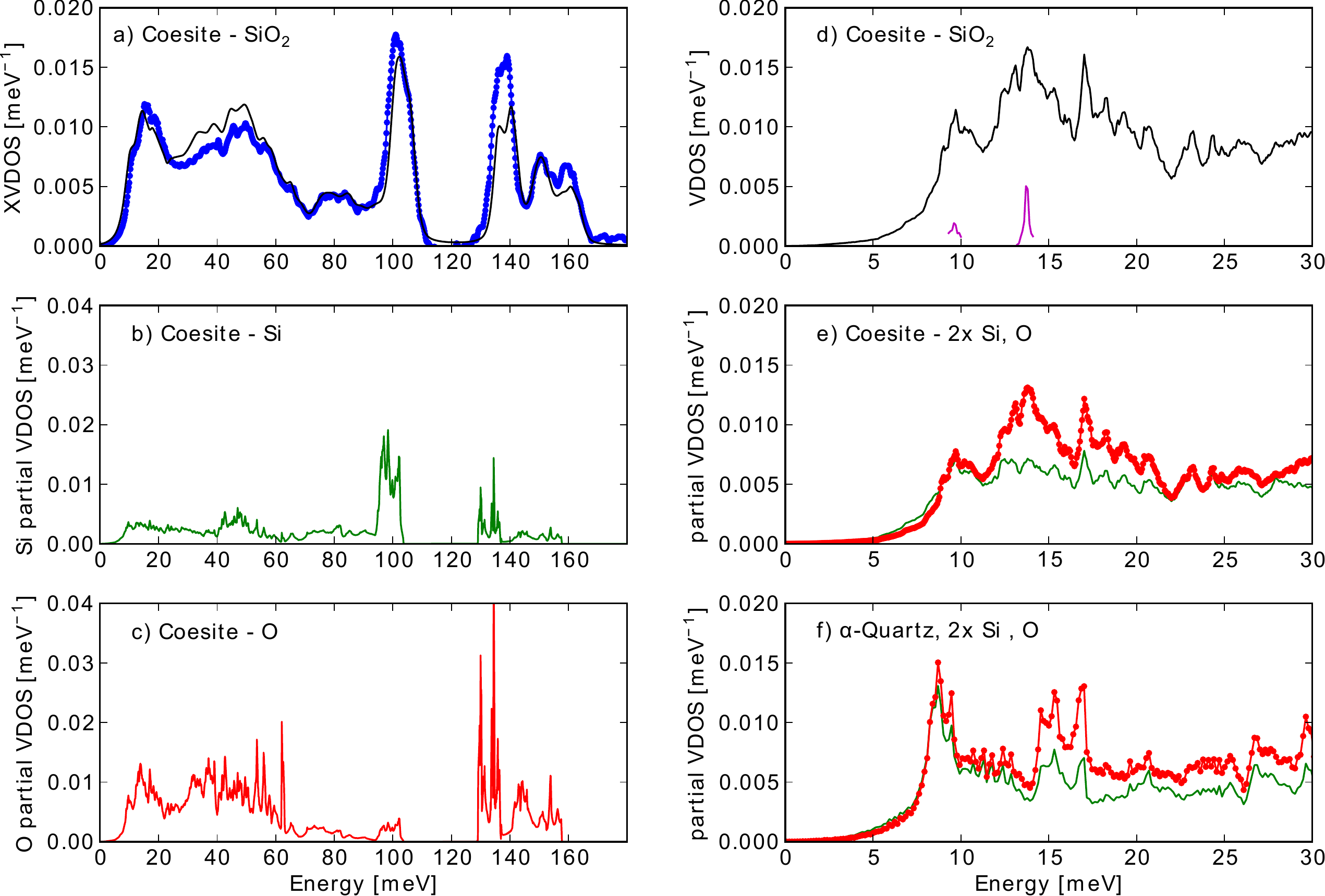}
\caption{\label{fig:VDOS}(colour online)  a) Experimental (blue points) and theoretical (black line) X-VDOS of coesite. The calculated X-VDOS is the x-ray weighted sum of the partial VDOS of oxygen (b) and silicon (c). The energies of the theoretical data were scaled by 1.045 and the resulting spectrum was broadened by the experimental resolution. d) Low-energy part of the VDOS (black line) and the local contribution of the critical points to the VDOS within $\mathbf{q}_c \pm 0.1 r.l.u.$ (purple line). The partial VDOS of oxygen (red dots) and silicon (green line) are compared in (e), where the silica contribution is multiplied by a factor of two. f) The partial VDOS of oxygen (red dots) and silicon (green line) in the low-energy range of $\alpha$-quartz. The silica contribution is multiplied by a factor of two.}
\end{figure}

The generalized x-ray weighted VDOS (X-VDOS),
\begin{equation}
G(E) = \frac{1}{N} \int d^3\bi{Q} \sum_j \vert \sum_n \frac{f_n(\bi{Q})}{\sqrt{m_n}}e^{i \bi{Qr}_n-W_n} (\bi{Q}\cdot\hat{\bi{e}}_{\bi{Q},j,n}) \vert^2 \delta[E-E_{\bi{Q},j)}] ,
\end{equation}

with momentum transfer $\bi{Q}$, phonon energy $E$ and eigenvector component $\hat{\bi{e}}$ of branch $j$, atomic form factor $f$, mass $m$, position vector $\bi{r}$, Debye Waller factor $W$ of atom $n$ and normalization factor $ N = \sum_n f_n(\bi{Q})/\sqrt{m_n} $ for coesite is shown in Figure \ref{fig:VDOS} a). The X-VDOS was used for the determination of an overall energy scaling factor. In fact the VDOS probes the energy of the ensemble of states in three dimensional reciprocal space and is therefore most significant for the determination of the scaling factor. It turns out that an overall stretching of 1.045 leads to a good agreement of experimental and theoretical X-VDOS. Discrepancies between experiment and theory are mostly due to the limited accuracy of sampling the reciprocal space with powder IXS spectra.
The underestimation of the calculated energies can be attributed to the limited accuracy of the exchange correlation function within the local density approximation; see Ref. \cite{refson_prb_2006} for a detailed discussion. The partial density of states (Figure \ref{fig:VDOS} b and c)) separate the contribution of silicon and oxygen atoms. Looking at the low energy part of the partial VDOS (Figure \ref{fig:VDOS} e)) we find that the first peak located at 10.1 meV has equal contributions from silicon and oxygen atoms. The main peak of the scattering function at 14.4 meV is, however, dominated by the vibration of the oxygen atoms. The low energy VDOS of $\alpha$-quartz is plotted in Figure \ref{fig:VDOS} f) for comparison. The $\alpha$-quartz calculation was performed using the same pseudopotentials and similar parameters as for coesite. For a detailed discussion on the calculation see Ref. \cite{refson_prb_2006}. Here the first peak dominates the low energy part of the VDOS. It arises as in coesite from equal contributions from silicon and oxygen atoms.

\begin{figure}[tb]
\includegraphics[width=0.8\textwidth]{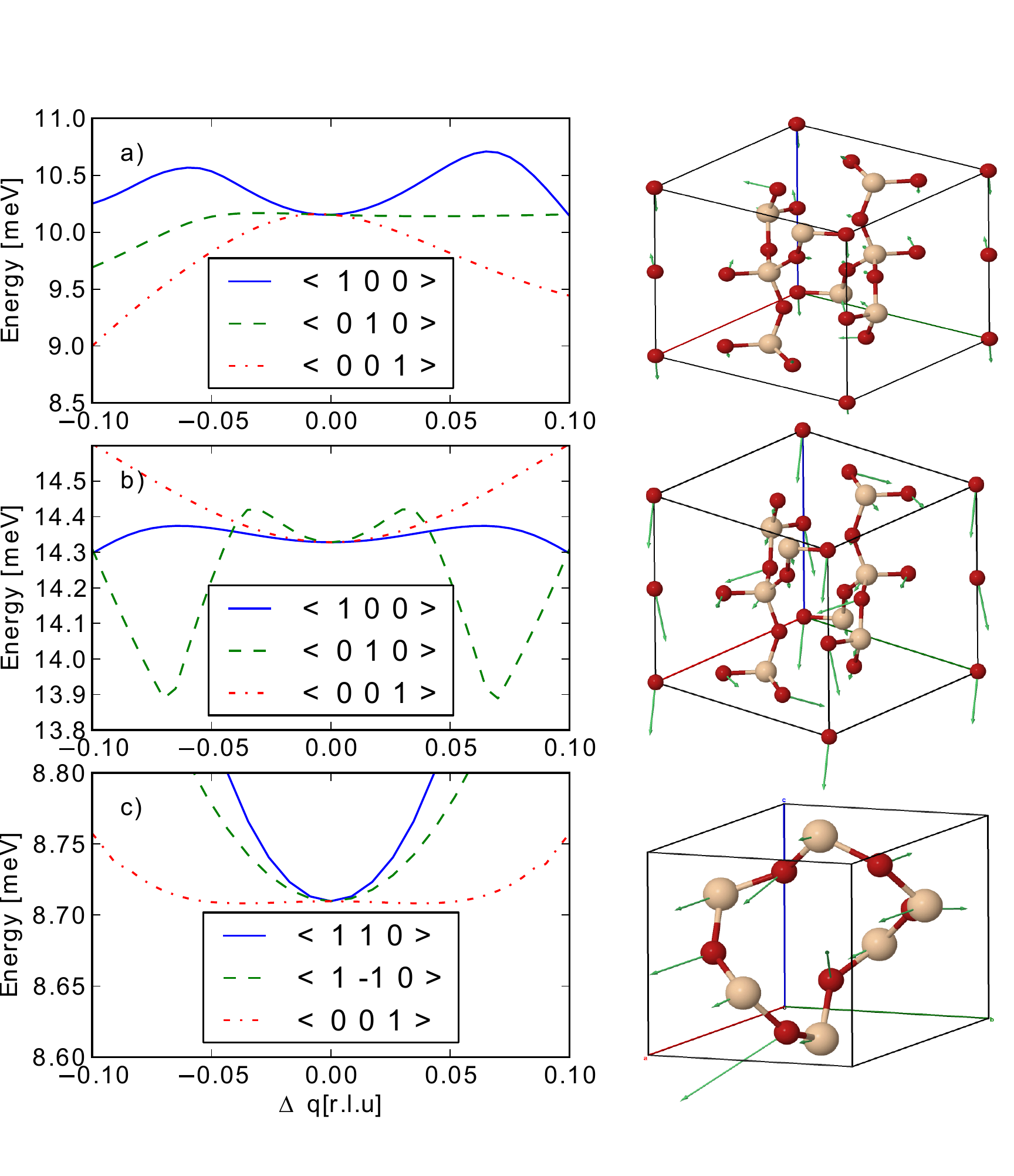}
\caption{\label{fig:CP} (colour online) a) Dispersion relations along the main crystallographic directions through the critical point (0.87 0.69 0.42) and the displacement pattern (primitive cell) of the atoms at this saddle point contributing strongly to the first peak in the VDOS of coesite. b) Dispersion relations and displacement pattern at the Y point (1 0 0) in coesite. c) Orthogonal dispersion relations and displacement pattern at the L point (1/2 0 1/2) in $\alpha$-quartz.}
\end{figure}

Peaks in the VDOS require critical points in the phonon dispersion relations for which the gradient in all crystallographic directions becomes zero \cite{vanhove_pr_1953}. In order to localize the critical points contributing most to the first and the main peak in the low energy part of the VDOS, two filters were applied simultaneously. An energy filter of $\Delta \omega $= 0.3 meV was applied to the \textit{ab initio} calculated phonon eigenfrequencies of the first Brillouin zone and $1/|\nabla_{\mathbf{q}}\omega(\mathbf{q})|$ was computed within this energy window. A saddle point close to the zone boundary at (0.87 0.69 0.42) was found to contribute strongly to the first peak in the VDOS and a local minimum at Y (1 0 0) is responsible for the main peak of the VDOS. The local contribution of the critical points to the VDOS within $\mathbf{q}_c \pm 0.1 r.l.u.$ is shown in Fig. \ref{fig:VDOS} d). The dispersion relations along the main crystallographic direction through the critical points and the displacement patterns are shown in Figure \ref{fig:CP} and compared to the ones of $\alpha$-quartz.  

Summarizing the results shown in Figure \ref{fig:VDOS} and \ref{fig:CP} we note: (i) Critical points close to or at the zone boundary contribute most to the first and main peak of the low energy VDOS in both systems. This observation might be explained in a simplified picture with the piling up of vibrational states due to a flattening of the dispersion relations at the zone boundary \cite{vanhove_pr_1953}. (ii) The first peaks in both coesite and in $\alpha$-quartz are due to an almost equal contribution from silicon and oxygen atoms. The main peak in coesite is, however, dominated by the vibration of oxygen atoms. The character of vibration is more libration like. (iii) In coesite both the first and the main peak are located at higher energies than in $\alpha$-quartz. This observation could be expected from the higher density structure, but we observe that (iv) the atomic displacements of both critical points in coesite are different from the atomic displacement pattern of $\alpha$-quartz. This shows that the peaks cannot be compared directly.

\begin{figure}[tb]
\includegraphics[width=0.8\textwidth]{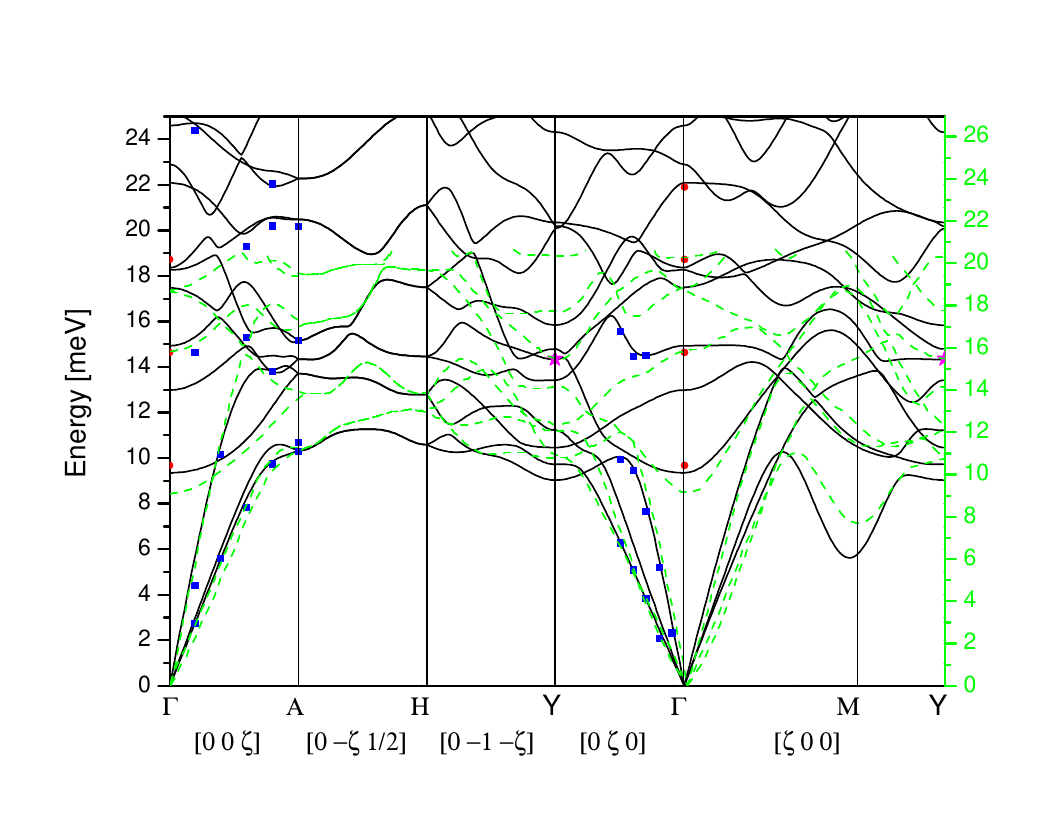}
\caption{\label{fig:dispersion} (colour online) Calculated dispersion relations along high symmetry directions of coesite crystal (black lines) and the results of an interatomic potential calculation \cite{dean_prb_2000} (dashed green lines) together with experimental values from the IXS measurements (blue squares) and Raman measurements \cite{liu_pcm_1997} (red points). The critical point at Y is marked with a magenta star. The energies of our calculation are scaled by 1.045 whereas the energies of interatomic potential calculation are scaled arbitrary for best visual fit.}
\end{figure}

Calculated dispersion relations along high symmetry directions and the result of an interatomic potential calculation \cite{dean_prb_2000} together with experimental values from the IXS measurements and Raman scattering \cite{liu_pcm_1997} are shown in Figure \ref{fig:dispersion}. The two calculations are in reasonable qualitative agreement for most of the branches. For the Y - $\Gamma $ - direction our calculation agrees much better with the experimental results, particularly for the fast acoustic and the measured optic mode. The calculation of the low energy optical branches is very delicate and sensitive to small geometrical differences and require a fine electronic grid sampling. We note that both acoustic and optical phonon branches contribute to the main peak of the scattering function at 14.4 meV. The M - Y dispersion relation of the branch containing the critical point is very flat. Experimental phonon energies as determined by the IXS measurement and Raman scattering \cite{liu_pcm_1997} are in good agreement with our calculation.

The calculation is further compared to experimental results from Raman and infrared \cite{lippincott_jrnbs_1958} studies in Table \ref{tab:RamanIr}. The modes were attributed by a careful comparison of experimental and calculated intensities. A good overall accordance is obtained. The calculation predicts some additional modes to be Raman or infrared active with low intensity contribution.  

\section{Conclusions}
With the validated lattice dynamics calculation it is possible to access the dynamical properties at arbitrary momentum transfer. The investigation of the VDOS, probing the ensemble of vibrational states, allows the identification of distinct dynamical features. In the case of coesite a small linear energy scaling of the ab initio calculated phonon frequencies leads to a good agreement of experiment and theory.
The investigation of the nature of the dominating features in the low energy part of the VDOS shows that critical points located close to or at the zone boundary provide the largest contribution. The contribution of oxygen atoms is found to dominate the main peak. The extension of the model for the lattice dynamics at ambient conditions to high pressure potentially builds the basis for understanding the compression mechanism and phase stability. This extension may provide an accurate model for the elastic properties at the formation of coesite inclusions in diamond implying a precise calibration of the coesite-in-diamond barometer proposed in Ref. \cite{sobolev_pnas_2000}. Comparison with the most abundant silica polymorph $\alpha$-quartz reveals valuable new insights into the low-energy vibrational properties of this prototypical oxide. Within the framework of silica polymorphs we are now able to provide accurate models of the lattice dynamics of $\alpha$-quartz \cite{bosak_zkri_2012}, coesite and stishovite \cite{bosak_grl_2009}. The calculation for these polymorphs can be extended to high pressures and allow the derivation of thermodynamical properties. The prediction of the lattice dynamics for new high pressure phases like seifertite and others \cite{dubrovinsky_pepi_2004} should be possible within the employed calculation scheme.

\section*{References}
 \bibliographystyle{unsrturl}
\bibliography{references}

\clearpage

\begin{table}[tb]
\caption{Cell parameters of SiO$_2$ coesite.} 
\label{tab:cell}
\begin{tabular}{l | l }
  Calculation                               &  Experiment \cite{smyth_jpc_1987} \\
  \hline
  a =  \hspace{2pt} 7.137 \AA \hspace{1pt}  &  a = \hspace{2pt} 7.136 \AA \hspace{1pt} \\
  b = 12.295 \AA \hspace{1pt}               &  b = 12.384 \AA \hspace{1pt} \\
  c =  \hspace{2pt} 7.072 \AA \hspace{1pt}  &  c = \hspace{2pt} 7.186 \AA \hspace{1pt} \\
  $\alpha$ = $\gamma$ = 90$^{\circ}$        &  $\alpha$ = $\gamma$ = 90$^{\circ}$ \\
  $\beta$  = 120.374$^{\circ}$              &  $\beta$ = 120.375$^{\circ}$  \\

\end{tabular}
\end{table}

\begin{table}[tb]
\caption{Raman and infrared energies [meV] of coesite at ambient conditions. The calculated values are scaled by 1.045. Intensity contributions are indicated by very weak (vw), weak (w), strong (s) and very strong (vs).} 
\label{tab:RamanIr}
\begin{tabular}{r r r r | r r r r }
  \multicolumn{4}{c|}{Raman} & \multicolumn{4}{c}{infrared} \\
  Calculation & &  Experiment \cite{liu_pcm_1997} & & Calculation & & Experiment \cite{lippincott_jrnbs_1958} & \\
  \hline
  9.4 &  w &   9.7 &  s &  13.0 & vw &     - &   \\
 14.9 &  s &  14.6 &  s &  17.5 & vw &     - &   \\
 18.3 &  w &  18.7 &  w &  18.4 & vw &     - &   \\
 22.1 &  s &  21.9 &  s &  22.9 & vw &     - &   \\
 24.6 & vw &     - &    &  31.2 & vw &     - &   \\
 25.1 &  s &  25.3 &  w &  32.5 & vw &     - &   \\
 30.3 &  w &  30.3 &  w &  33.8 & vw &     - &   \\
 33.8 &  s &  33.6 &  s &  36.4 & vw &     - &   \\
 35.9 & vw &     - &    &  36.9 & vw &     - &   \\
 39.3 &  w &     - &    &  37.4 & vw &     - &   \\
 40.9 &  w &  40.4 &  w &  40.7 & vw &     - &   \\
 44.2 &  w &  44.1 &  w &  42.2 &  w &  42.2 &  w\\
 47.2 &  w &  47.0 &  w &  46.2 &  w &     - &   \\
 53.1 &  s &  52.9 &  w &  46.8 &  w &     - &   \\
 55.1 &  w &     - &    &  48.9 &  w &  48.3 &  w\\
 57.6 & vw &     - &    &  52.1 &  w &  53.3 &  w\\
 58.5 &  w &  57.8 &  w &  52.2 &  w &     - &   \\ 
 65.7 & vs &  64.6 & vs &  55.0 &  w &  54.8 &  w\\
 68.4 &  w &     - &    &  56.6 &  w &     - &   \\
 84.2 & vw &     - &    &  61.4 &  w &     - &   \\
 85.6 &  w &     - &    &  69.6 &  w &  69.1 &  w\\
101.3 &  w &  97.6 &  w &  75.9 &  w &  74.1 &  w\\
101.6 &  s & 101.2 &  w &  85.8 & vw &     - &   \\
105.1 &  s & 104.0 &  w &  87.1 &  w &  84.7 &  w\\
105.9 & vw & 105.5 &  w & 102.5 &  w &  98.7 &  w\\
108.1 &  w & 110.7 &  w & 105.0 &  w & 100.8 &  w\\
135.6 &  s & 132.2 &  w & 108.1 & vw &     - &   \\
136.1 &  w & 132.8 &  w & 135.4 & vs & 129.0 & vs\\
139.7 &  w & 139.0 &  w & 139.5 & vs & 136.1 & vs\\
140.5 &  w & 141.8 &  w & 140.4 &  w &     - &   \\
149.1 &  s &     - &    & 142.8 &  s & 145.1 &  w\\
151.0 &  s &     - &    & 149.3 &  w &     - &   \\
153.2 &  w &     - &    & 150.4 &  w & 151.9 &  w\\ 
      &    &       &    & 159.3 & vw &     - &   \\
      &    &       &    & 160.6 &  w &     - &   \\
\end{tabular}

\end{table}

\end{document}